\documentclass[draft,grl]{agutex}

\usepackage{amsmath,graphicx,amssymb}

 \setkeys{Gin}{draft=false}
\authorrunninghead{Jougnot et al.}

\titlerunninghead{Seismoelectrics \& mesoscopic effects}

\authoraddr{D. Jougnot,
Applied and Environmental Geophysics Group, University of Lausanne, CH-1015 Lausanne, Switzerland.
(Damien.Jougnot@unil.ch)}

\graphicspath{{figures/}}

\begin{document}

%
%

\title{Seismoelectric effects due to mesoscopic heterogeneities}
%

%
%


\author{Damien Jougnot$^*$} 

\author{J. Germ\'an Rubino$^*$} 

\author{Marina Rosas Carbajal} 

\author{Niklas Linde} 

\author{Klaus Holliger} 
\affil{Applied and Environmental Geophysics Group, University of Lausanne, CH-1015 Lausanne, Switzerland.}

\let\thefootnote\relax\footnotetext{$^*$ These authors contributed equally to this work}

\begin{abstract}

\noindent
While the seismic effects of wave-induced fluid flow due to mesoscopic
heterogeneities have been studied for several decades,
the role played by these types of heterogeneities on seismoelectric
phenomena is largely unexplored. To address this issue, we have developed a novel
methodological framework which allows for the coupling of wave-induced fluid flow,
as inferred through numerical oscillatory compressibility tests,
with the pertinent seismoelectric conversion mechanisms. 
Simulating the corresponding response of a water-saturated sandstone sample containing
mesoscopic fractures, we demonstrate for the first time that these kinds of heterogeneities can produce 
measurable seismoelectric signals under typical laboratory conditions.
Given that this phenomenon is sensitive to key hydraulic and mechanical properties,
we expect that the results of this pilot study will stimulate
further exploration on this topic  in several domains of the  Earth, environmental, and
engineering sciences.

\end{abstract}

%
%

%

\begin{article}
%
%

\section{Introduction}

When a seismic wave propagates through a water-saturated porous medium, it produces a 
relative motion between the fluid phase and the rock matrix \citep{biot_low}. 
This flow transports an electrical excess charge, contained in the electrical double layer
located along grain surfaces, thus producing an electrical current source.
This is the physical principle of 
the seismolectric phenomenon, which \cite{pride94} formalized by coupling 
Biot's (\citeyear{biot_low}) and Maxwell's equations. 

For seismic waves propagating through water-saturated porous media,
 the theory predicts two kinds of seismoelectric conversions:
(1) the coseismic field, and (2) the interface response.
The coseismic field is a consequence of the wavelength-scale flow accompanying a compressional 
wave, which generates a current source even in homogeneous media.
The resulting electric field travels with the wave
and has a very limited extent outside of the support of the wave. Conversely, 
when a seismic wave encounters a contrast in mechanical or electrical
properties, there is a corresponding variation in the current source 
distribution, thus  generating electrical potential differences that  
can be measured outside of the wave support. The associated electric fields are highly sensitive 
to the fluid pressure gradients in the vicinity of the interface. Accurate  
modeling of seismic wave conversions at interfaces and, in particular,
of the associated Biot slow waves, is therefore critical for a realistic simulation of 
the seismolectric response \citep{pride2002biot}.

A particular interface-type response is expected to take place in the presence of mesoscopic
heterogeneities, that is, heterogeneities having sizes larger than the 
characteristic pore scale but smaller than the prevailing wavelengths. 
It is well known that the propagation of seismic waves through a medium containing 
these kinds of heterogeneities can induce significant 
oscillatory fluid flow as, in response to the spatial variations in
elastic compliances, the stresses associated with the passing seismic wave produce a pore 
fluid pressure gradient.
Indeed, the energy dissipation 
associated with this phenomenon is widely considered to be one of the
most important intrinsic seismic 
attenuation mechanisms in the shallower parts of the crust \citep[e.g.,][]{muller-et-al10}.
 The characteristics of such wave-induced flow are mainly controlled by the compressibility contrast
between the heterogeneities and the embedding matrix as well as permeabilities and 
geometrical characteristics of the heterogeneities.
Given that the amount of flow produced by this phenomenon can 
be significant, corresponding strong seismoelectric signals carrying 
valuable information about these properties are also expected to arise.

Modeling wave-induced fluid flow is problematic because the corresponding diffusion
lengths, that is, the spatial scales at which fluid flow occurs,
are very small compared with the seismic wavelengths. Together with the theoretical complications
arising from the coupling of the poro-elastic and electromagnetic responses, this may
explain why the generation of seismoelectric effects due to  
mesoscopic heterogeneities is largely unexplored. Arguably, the most important work
on this topic is by \cite{haartsen1997Electroseismic}, who modeled the seismoelectric
response from a single sand layer having a thickness much smaller
 than the predominant seismic wavelengths.
More recently, \cite{zhu2008electroseismic} 
performed seismoelectric laboratory experiments demonstrating that electromagnetic waves are generated at horizontal fractures intersecting boreholes.

In this paper, we propose a novel approach to address this problem based on a numerical oscillatory 
compressibility test coupled with a model for seismoelectric 
conversion and signal generation.  We illustrate the methodology
by analyzing the electrical potential produced by mesoscopic 
fractures in an otherwise homogeneous water-saturated sandstone sample.
The reason for this choice of model is that 
the amount of wave-induced fluid flow scales with the compressibility
contrasts between the mesoscopic heterogeneities and their embedding 
matrix, which in turn implies that a particularly prominent seismoelectric 
response can be expected in fractured media.

\section{Methodological background}

We consider a square rock sample containing mesoscopic heterogeneities and
apply  a time-harmonic compression 
\begin{equation}\label{source}
P(t)=\Delta P\cos{\left(\omega t\right)},
\end{equation} at its top boundary,
where $\omega$ is the angular frequency and $t$ time. 
No tangential forces are applied on the boundaries of the sample 
and the solid is neither allowed to move on the bottom boundary nor to 
have horizontal displacements on the lateral boundaries.
The fluid is not allowed to flow into or out of the sample. 
To obtain the response of the sample, 
we numerically solve the equations of quasi-static poroelasticity 
under corresponding boundary conditions \citep{rubino-et-al09}. This 
methodology allows for 
computing the relative fluid velocity field $\dot{\boldsymbol w}$ which 
is then employed to calculate the corresponding seismoelectric signal.

Wave-induced flow exerts a drag on the excess electrical charges of the double layer
surrounding grain surfaces, thereby generating a source current density of the 
form \citep{jardani2010stochastic}       
\begin{equation}
	\textbf{j}_s = \bar{Q}_v^{eff} \dot{\boldsymbol w},
	\label{eq:Js}
\end{equation}
where $\bar{Q}_v^{eff}$ is the effective excess charge density.

In absence of an external current density, the electrical potential  $\varphi$ in 
response to a given source current density is described by \citep{Sill1983SP}
\begin{equation}
	\nabla \cdot (\sigma \nabla \varphi) = \nabla \cdot \textbf{j}_s,
	\label{eq:poisson}
\end{equation}
where $\sigma$ denotes the electrical conductivity. Given  the 
fluid velocity field $\dot{\boldsymbol w}$ inferred from the oscillatory compressibility test,
the seismoelectric signal induced by the presence of mesoscopic heterogeneities
is then obtained by numerically solving equations (\ref{eq:Js}) and (\ref{eq:poisson}). 
As proposed by \cite{revil2013coupled}, we choose to ignore electroosmotic phenomena.
Conversely, we assumed the electrical conductivity to be invariant with frequency 
because the frequency-dependence of the effects of the wave-induced fluid flow
are orders-of-magnitudes larger than those of the electrical conductivity
(see \cite{kruschwitz2010textural} for electrical measurements on sandstones).

\section{Numerical example: Seismoelectric response of fractured rocks}

\begin{figure}\vskip -3cm
\centering\includegraphics[angle=0,width=0.90\textwidth]{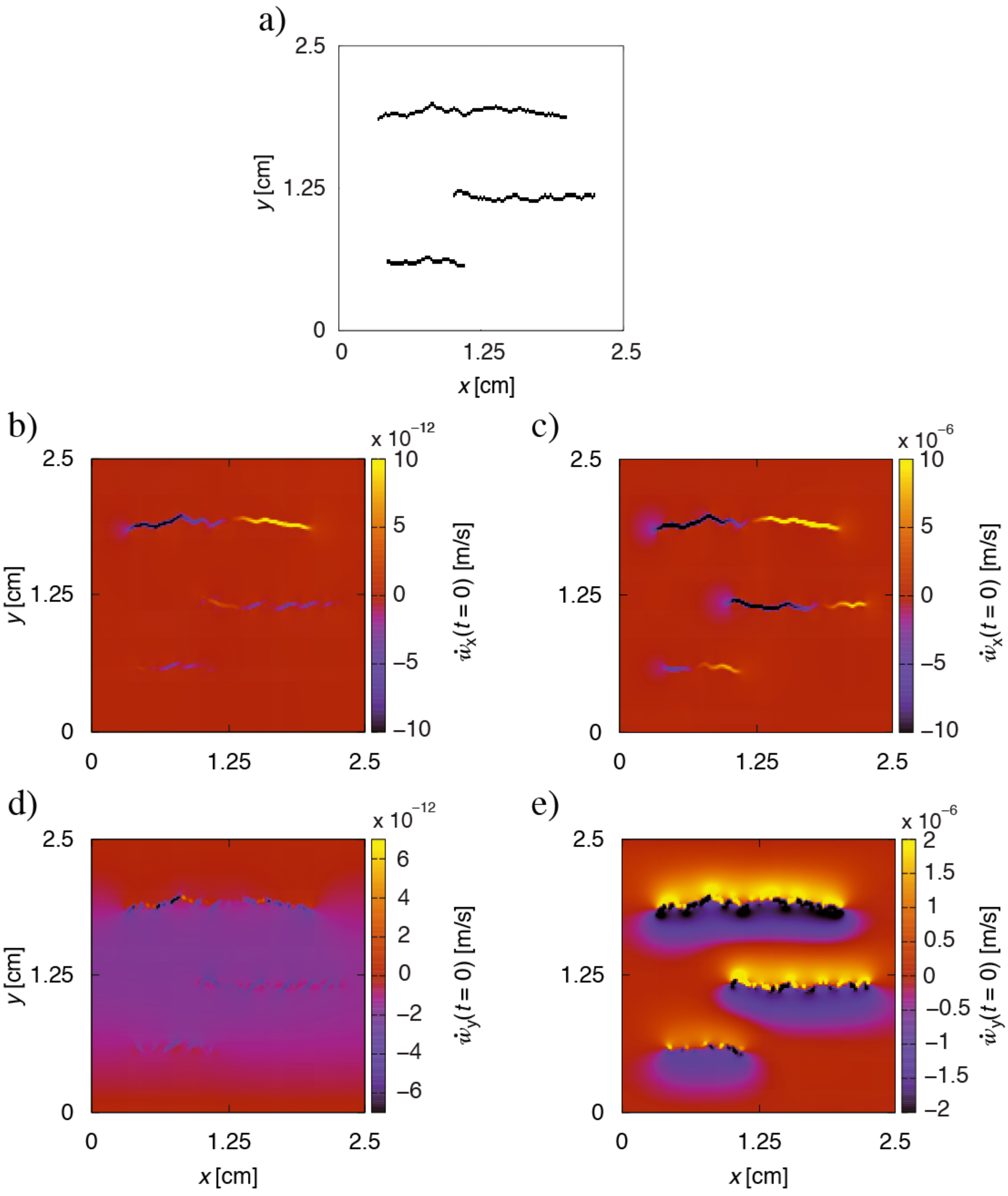}
\caption{(a) Numerical sample with three fractures.
(b and c) Horizontal and (d and e) vertical components of the 
relative fluid velocity field at $t=0$ in equation (\ref{source}). The results correspond to an
oscillatory compressibility test with an amplitude of 1~kPa and frequencies of
(b and d) 1~Hz and (c and e) 10~kHz.}
\label{fig:Geom&Vf}
\end{figure}

We consider a model of a homogeneous water-saturated clean sandstone
permeated by three mesoscopic fractures (Figure \ref{fig:Geom&Vf}a).
The sample is a square with a side length of $2.5$~cm, and the mean 
aperture of the fractures  is $h=0.033$~cm. 
The poroelastic response of this medium is modeled by
representing the mesoscopic fractures as highly compliant and
permeable porous regions embedded in a 
stiffer porous matrix \citep{rubino2013fracture}. 
The background and the fracture materials are hereafter denoted through the 
superscripts \textit{b} and \textit{f} \citep{rubino2013fracture}:  
drained-frame bulk moduli $K_m^b =$ 23~GPa and $K_m^f =$ 0.02~GPa,
shear moduli $\mu_m^b =$ 27 GPa and $\mu_m^f =$ 0.01~GPa,
porosities $\phi^b =$ 0.1 and $\phi^f =$ 0.5, 
permeabilities $\kappa^b =$ 2.37 $\times$ 10$^{-14}$~m$^2$ and $\kappa^f =$ 10$^{-10}$~m$^2$,
and solid grain bulk moduli $K_s^b=K_s^f=37$~GPa.
The sample is fully saturated with water, with bulk modulus $K_w =$ 2.25~GPa and
viscosity $\eta_w =$ 0.003~Pa s. The amplitude of the applied compression is $\Delta P=1$~kPa.

We determine the electrical conductivities using $\sigma = \sigma_w \phi^{m}$, 
with $m$ denoting the cementation exponent from Archie's law 
and $\sigma_w$ the pore water conductivity.
Given the considered medium, the surface conductivity can be neglected.
We use $m^b$=2, $m^f$=1.3, and $\sigma_w$ = 10$^{-2}$~S/m, which represents 
a typical value for pore water conductivity in laboratory experiments. 

The simulated flow is in the viscous laminar regime and we can therefore estimate
the  effective excess charge for the two materials using the empirical relationship proposed by 
\cite{jardani2007tomography}
\begin{equation}
	\log{(\bar{Q}_v^{eff})} = - 9.2349 - 0.8219 \log{(\kappa)}.
	\label{eq:Qveff}
\end{equation}
This yields $\bar{Q}_v^{eff,b}$=87.6~C/m$^{3}$ and 
$\bar{Q}_v^{eff,f}$=9.33~$\times 10^{-2}$~C/m$^{3}$. These values are consistent with those obtained by \cite{jouniaux1995permeability}
 from voltage coupling coefficient measurements on sandstones with similar permeabilities.

Figure \ref{fig:Geom&Vf} displays the horizontal ($x$) and vertical ($y$) components
of the relative fluid velocity field at $t=0$ in equation (\ref{source}). For 1~Hz (Figures \ref{fig:Geom&Vf}b 
and \ref{fig:Geom&Vf}d), we observe that the induced fluid velocity field is negligible.
This is expected, as for such a low frequency the diffusion lengths are larger than the size of 
the considered heterogeneities and there is enough time during each oscillatory
half-cycle for the pore fluid pressure to equilibrate to a common value.
For a frequency of 10~kHz, the oscillatory compression produces 
a significant fluid pressure increase in the highly compliant 
fractures as compared to the stiffer embedding matrix, thus establishing
an important fluid pressure gradient.
Correspondingly, the amount of fluid flow is much more important in this case
(Figures \ref{fig:Geom&Vf}c and \ref{fig:Geom&Vf}e). Significant fluid flow
occurs inside the fractures (Figure \ref{fig:Geom&Vf}c), but there is also 
an important fluid exchange between the fractures and the embedding matrix material 
(Figure \ref{fig:Geom&Vf}e).

\begin{figure}
\centering\includegraphics[angle=0,width=1\textwidth]{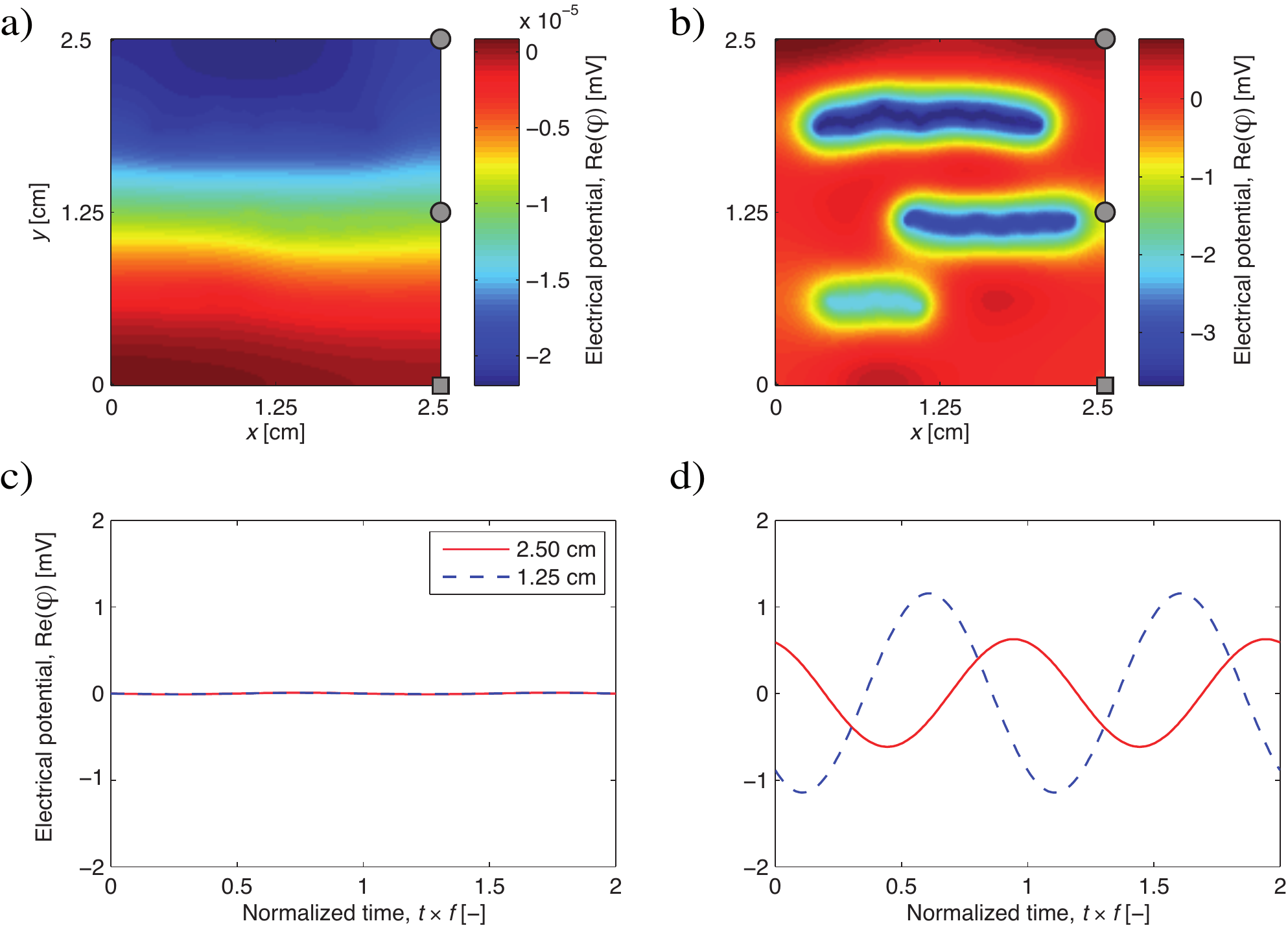}
\caption{(a and b) Electrical potential distribution at $t=0$ in equation (\ref{source}) and (c and d)
resulting  electrical potential differences at two electrodes with respect to 
the reference electrode as functions of the normalized time.
The grey square and the two circles (in a and b) highlight the position of the reference 
and the potential electrodes, respectively. 
The results correspond to an oscillatory compressibility test with an amplitude of 1~kPa
and frequencies of (a and c) 1~Hz and (b and d) 10~kHz.}
\label{fig:Voltage}
\end{figure}

The relative fluid velocity field obtained from the oscillatory compressibility test 
is used to solve the electrical problem through equations (\ref{eq:Js}) and (\ref{eq:poisson}). 
Neumann boundary conditions (electrical insulation) are considered at the boundaries of 
the sample, in conjunction with a Dirichlet boundary condition ($\varphi=0$~V) at the right bottom corner
of the sample ($x=$2.5 and $y=$0~cm).

Figures \ref{fig:Voltage}a and \ref{fig:Voltage}b show the 
electrical potential at $t=0$ for 1~Hz and 10~kHz, respectively. The magnitude and distribution
of the seismoelectric signal strongly depends on the frequency of the applied compression.
In particular, the fluid flow occurring in the vicinity
of the fractures induces at 10~kHz  an important divergence of the source current density 
$\textbf{j}_s$, which in turn generates seismoelectric signals with measurable amplitudes of a few mV
(Figure \ref{fig:Voltage}b).
For a frequency of 1~Hz, the signal is too small to be measured (Figure \ref{fig:Voltage}a),
which is expected given the smaller magnitude of fluid flow (Figures \ref{fig:Geom&Vf}b and \ref{fig:Geom&Vf}d).
 
To evaluate the feasibility of observing such seismoelectric
signals in laboratory experiments,
we consider the responses at two measurement electrodes with respect to a reference electrode
($\varphi=0$~V) as functions of ``normalized'' time ($t \times f$)
(Figures \ref{fig:Voltage}c and \ref{fig:Voltage}d).
While no clear signal can be seen for a frequency of 1~Hz (Figure \ref{fig:Voltage}c),
the seismoelectric responses are on the order of a few mV  at 10~kHz 
(Figure \ref{fig:Voltage}d), which can be readily measured under laboratory conditions.
We also observe a discrepancy, both in magnitude $\vert \varphi \vert$ and 
phase $\theta$, between the two electrodes simulated for this higher frequency. 
Indeed, even though the electrode located at the top boundary is 
further from the reference than the one located in the middle,
its signal is 1.8 times weaker. This electrode is almost in phase 
($\theta\approx 20^\circ$) with the oscillatory compression, while the middle electrode  
shows a more important phase shift ($\theta\approx 140^\circ$). 
This phase shift is due to both the viscosity-related lag experienced by 
the wave-induced fluid flow and the relative position of the electrode with respect to the fractures.

\begin{figure}
\centering\includegraphics[angle=0,width=0.5\textwidth]{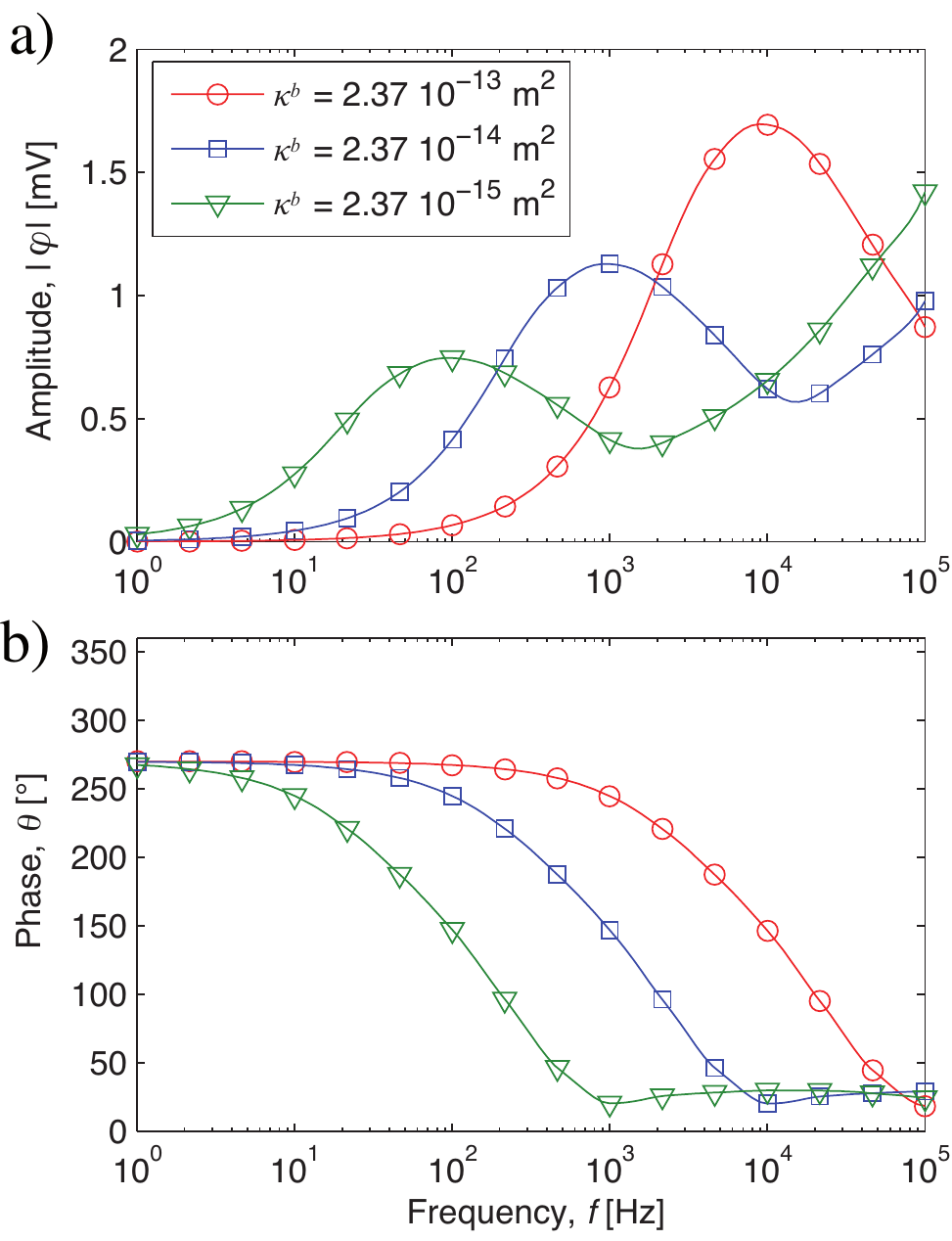}
\caption{Effect of background permeability upon  (a) the amplitude 
$\vert \varphi \vert$ and (b) the phase $\theta$ of the electrical voltage
recorded at the top electrode ($y$ = 2.5 cm) for different frequencies.}
\label{fig:Kappa_test}
\end{figure}

Additional tests were conducted for a wide frequency range using different background permeabilities. 
Figures \ref{fig:Kappa_test}a and \ref{fig:Kappa_test}b show that the electrical 
potential measured at the top electrode is strongly frequency-dependent in terms of its amplitude and 
phase. The amplitude has a first peak at a frequency that depends on the background 
permeability, followed by a general increase at higher frequencies (Figure \ref{fig:Kappa_test}a).
The phase spectrum also shows a dependence on $\kappa^b$,
as the transition from high to low phase angles $\theta$ shifts to lower 
frequencies as the background permeability decreases (Figure \ref{fig:Kappa_test}b).
These spectral signatures are explained by the
fact that the frequency range where fluid flow prevails scales with the background
permeability, together with the effects produced by the variations of the effective excess charge 
according to equation (\ref{eq:Qveff}).

\section{Discussion}

When a seismic wave is incident perpendicular to a mesoscopic fracture plane,
or equivalently, when an oscillatory compression is applied to a rock sample
containing sub-horizontal fractures,
an oscillatory flow is induced from the fracture into the pore space of the embedding matrix
and vice versa. The spatial scales at which this flow occurs is limited
by the spacing between the fractures and characterized by the diffusion length $L_{\rm d}\equiv\sqrt{D/\omega}$,
where $D$ is the pressure diffusivity of the embedding matrix, a 
parameter directly proportional to the background permeability. Therefore,
the frequency range where significant flow prevails scales with this hydraulic parameter and 
the spacing between the fractures. The amount of fluid flow is mainly governed by the compressibility contrast
between the fractures and the embedding matrix.
Therefore, while the frequency dependence of the seismoelectric signal is mainly governed
by the separation between fractures and the embedding matrix permeability, its magnitude is expected to be
controlled by the compressibility contrast between the fractures and the background,
in addition to the permeability of the background, which also affects the effective excess charge (equation \ref{eq:Qveff}).

Additional numerical simulations indicate that the seismoelectric
signal is highly sensitive to the orientation of the fractures. This is expected, as
the amount of fluid flow in response to a vertical compression is maximum for sub-horizontal fractures
and minimum for the sub-vertical case \citep{rubino2013fracture}.
Moreover, fracture connectivity is also expected to play an important
 role in determining the characteristics of  
seismoelectric signals \citep{rubino2013fracture}. 
Consequently, the seismoelectric responses due to mesoscopic effects
are expected to contain key information on structural and hydraulic
properties of the rock samples.

The most relevant result of this work is that mesoscopic heterogeneities can produce
measurable seismoelectric signals in response to an oscillatory compression
(Figures \ref{fig:Voltage} and \ref{fig:Kappa_test}). Corresponding laboratory experiments are already conducted
for seismic purposes \citep[e.g.][]{batzle2006fluid},
and could be extended to seismoelectric measurements. Although our analysis was performed 
considering laboratory-size samples, the results of this study also
have corresponding implications for seismoelectric conversions at the field scale.
Due to ubiquitous fractal scaling laws, virtually all
geological formations contain mesoscopic heterogeneities and, therefore, 
seismic waves will produce seismoelectric signals as they travel through such heterogeneities.
Indeed, it is likely that some of the notorious difficulties encountered in seismoelectric field applications,
notably the generally high noise levels \citep[e.g.][]{strahser2011dependence}, 
could be related to heterogeneities of different nature and sizes,
yielding a multiplicity of seismoelectric source currents 
dispersed over the volume traversed by the seismic waves.
A better understanding of the role played by mesoscopic heterogeneities
is therefore needed to improve the generation, recording and interpretation of seismoelectric
signals.

\section{Conclusions}

Based on a novel methodological framework that couples recent improvements in the modeling
of wave-induced fluid flow and seismoelectric conversion mechanisms,
we have shown for the first time that the presence of mesoscopic heterogeneities
can produce measurable seismoelectric signals for typical laboratory
setups. In particular, we find a measurable
frequency-dependent response of the seismoelectric signal caused by mesoscopic fractures 
in an otherwise homogeneous water-saturated sandstone sample.  
The magnitude of the seismoelectric signal is mainly governed by the compressibility contrast 
between the embedding matrix and the mesoscopic heterogeneities. Therefore, 
prominent seismoelectric effects are expected to arise not only in fractured media,
but also in partially saturated porous rocks. 
This in turn opens the perspective of developing seismoelectric spectroscopy
 as a novel method for characterizing such media.

Given the ever increasing interest in the measurement and interpretation of seismoelectric signals,
the results of this pilot study are expected to be of interest in a wide range of domains
of the Earth, environmental, and engineering sciences, including 
nondestructive testing, groundwater and contaminant hydrology,
hydrothermal operations, nuclear waste storage as well as hydrocarbon exploration and production,
among many others.
Correspondingly, further computational and experimental work on this topic is needed, as it promises to provide
deeper insights on the dependence of the recorded signals on pertinent hydraulic and mechanical
properties as well as to improve the acquisition, recording, and interpretation of seismoelectric data
{\it per se}.

\begin{acknowledgments}
This work was supported in part by grants from the Swiss National Science Foundation. The authors modified Maflot (maflot.com), kindly provided by I. Lunati and R. Kunze. The authors thank the Editor Michael Wysession and two anonymous reviewers for constructive comments that helped to improve the quality of this manuscript.
\end{acknowledgments}

%
%


\end{article}

\end{document}